\documentclass[pre,onecolumn,amssymb]{revtex4}
\usepackage{epsfig,amssymb,amsmath}

\newcommand{\be}{\begin{equation}} 
\newcommand{\ee}{\end{equation}}

\begin{document}
\title{Force network analysis of jammed solids}
\author{S.-L.-Y. Xu$^1$, X. Illa$^{1*}$, and J. M. Schwarz$^1$}
\affiliation{$^1$Physics Department, Syracuse University, Syracuse, NY 13244}

\date{\today}

\begin{abstract}
Using a system of repulsive, soft particles as a model for a jammed solid, we analyze its force network as characterized by the magnitude of the contact force between two particles, the local contact angle subtended between three particles, and the local coordination number. In particular, we measure the local contact angle distribution as a function of the magnitude of the local contact force. We find the suppression of small contact angles for locally larger contact forces, suggesting the existence of chain-like correlations in the locally larger contact forces. We couple this information with a coordination number-spin state mapping to arrive at a Potts spin model with frustration and correlated disorder to draw a potential connection between jammed solids (no quenched disorder) and spin glasses (quenched disorder). We use this connection to measure chaos due to marginality in the jammed system. In addition, we present the replica solution of the one-dimensional, long-range Potts glass as a potential toy building block for a jammed solid, where a sea of weakly interacting spins provide for long-range interactions along a chain-like backbone of more strongly interacting spins.
\end{abstract}

\maketitle

\section{Introduction}
\label{sec:intro}
The jamming phase diagram depicts a plausible scenario for a unified description of the phase change/crossover from a liquid to an amorphous solid in
nonequilibrium systems ranging from granular particles to colloidal particles to molecular particles~\cite{liu,jammingbook}.  These phase changes can be driven by temperature changes, packing fraction changes, and/or changes in applied shear stress. Whether or not the boundaries of the jamming phase diagram are sharp in the equivalent equilibrium sense is still a matter of debate, particularly for molecular particles in the presence of temperature changes~\cite{glassbook}.  However, numerical studies of repulsive, soft particles at zero-temperature indicate the
potential of a transition in the equivalent thermodynamic sense as the packing fraction of the system is increased~\cite{ohern1,ohern2}.

The numerics of the zero-temperature repulsive, soft particle system suggest that the transition from liquid to amorphous solid is somewhat of an unusual
nature~\cite{ohern1,ohern2,durian}. For
example, the average coordination number jumps from zero to some finite value at the transition, followed by a square root increase as a function of distance from the transition, i.e. $\phi-\phi_c$,
where $\phi$ denotes the packing fraction and $\phi_c$, the critical packing fraction above which the system behaves as
a solid.  Moreover, the jump in the average
coordination number
only depends on the dimension of the system. This is reminiscent of the universality of the value of the jump in the spin-wave
stiffness in the two-dimensional Kosterlitz-Thouless transition~\cite{kosterlitz}. While there exists a jump in a possible order parameter at the transition,  the square root scaling demonstrates that the transition
is not the typical first-order transition with no diverging correlation
lengths. In fact, there are several diverging correlation lengths above and
below the transition.  For example, the fraction of jammed configurations as a function of $\phi$ becomes
increasingly sharper as the system size is increased. This sharpening can be tied to an increasing lengthscale below
which the system acts as uncorrelated subsystems each with their respective critical packing fractions.

As for the jammed phase itself, it exhibits various interesting properties.
While the scaling of the bulk modulus with packing fraction is the same as ordered elastic solids, the scaling of the shear
modulus with packing fraction shows an anomalous response~\cite{ohern1,ohern2}.  In addition, the
vibrational modes are of a ``swirly'' nature and appear to be quasi-localized as the jamming transition is approached from the solid side~\cite{zeravcic,silbert}.
Similar behavior has been observed in other amorphous solids~\cite{laird,fabian,widmer,desouza}.  Both of these observations may be related to the
non-affine displacements of the particles, which become more pronounced near the jamming transition. There is also an excess in the vibrational density of states beyond the Debye prediction at the transition. This excess has been likened to the Boson peak~\cite{zeller,ediger,nakayama} in glassy systems. 

There have been a number of approaches to understand some of the above properties, including a statistical mechanics approach~\cite{henkes1}, a field theoretic approach~\cite{henkes2}, and a 
phenomenological approach~\cite{wyart1,wyart2,wyart3}. The phenomenological approach examines the consequences of the isostatic nature of the
transition by establishing a lengthscale below which the system behaves as an isostatic solid and above which the system
behaves as an ordinary elastic solid. This lengthscale has been inferred from the numerical simulations by relating the frequency below which there exists an excess in the density of states (beyond the Debye prediction) to a lengthscale via a linear dispersion relation~\cite{silbert2}.  However, a direct measurement of the elasticity of the system via a local perturbation shows the system to
respond on average as an ordinary elastic solid at all lengthscales at the jamming transition~\cite{ellenbroek1,ellenbroek2}. On the other hand, the study of fluctuations in the
response of all contacts some radial distance from the local perturbation demonstrates a clear
crossover lengthscale where the root-mean-squared fluctuations go from one type of power-law dependence on radial distance to another type further away from the perturbation. This crossover lengthscale appears to scale with distance from the transition.

One would like to understand why averaged response of the system behaves as ordinary elastic solid at the
transition.  Or, conversely, why a potential isostatic lengthscale may be showing up only in fluctuations to a response in position space. In order to
begin to investigate this mystery, we numerically study the force network of a two-dimensional system of repulsive, soft particles as the jamming transition is approached from above. In particular, we present evidence for the existence of chain-like structures in the
locally large force bonds and argue that the particles surrounding the chain-like structures give rise to long-range
interactions along the chain-like structures. However, the long-range interactions can be averaged out by the
proliferation of these chain-like structures just as long-range interactions are typically screened out in polymeric
systems. In addition, we propose a mapping of the coordination number to a
spin state in order to potentially better understand some of the properties of the jammed solid from what will turn out
to be a spin glass perspective.  Therefore, we will be able to draw direct links between spin glasses and amorphous
solids.   Our mapping should further open the door for other amenable analytical
techniques used in spin glasses with quenched disorder for studying amorphous solids where the disorder is not
quenched.  We note that there exists very interesting work on a replica-inspired approach to the glass transition for hard spheres as
approached from the liquid side (as opposed to the solid side)~\cite{zamponi}.

The paper is organized as follows. Section II presents our numerical analysis of the force network in the jammed phase.
Section III establishes a connection of the jammed solid and spin glasses via analysis of the connectivity, or
coordination number.  Section IV goes beyond the numerical findings in Sections II and III to speculate on a plausible
spin glass scenario for some properties of the jammed phase.  Section V concludes the paper with a discussion of the relation of this work to preexisting work in the field.

\section{Force bonds and spatial correlations}

To study the force network of a two-dimensional jammed solid of repulsive, soft particles, jammed configurations are
generated using the algorithm introduced by O'Hern and collaborators~\cite{ohern1,ohern2}. More specifically, the system consists of 50:50 mixture of $N$ particles with a diameter, $\sigma$, ratio of 1.4.  The packing fraction sets the radii for a system of length unity. The particles
interact via the following two-body potential:
\begin{equation}
V(r_{ij})= \frac{\epsilon}{\lambda}(1-\frac{r_{ij}}{\sigma_{ij}})^\lambda\,\,\Theta(1-\frac{r_{ij}}{\sigma_{ij}}),
\end{equation}
where $r_{ij}$ is the distance between the centers of the two particles $i$ and $j$, $\sigma_{ij}=(\sigma_i+\sigma_j)/2$, and $\epsilon$ sets the energy scale for the system.
The particles are placed randomly in the system with periodic boundary conditions and the conjugate gradient algorithm
is invoked until the system reaches its nearest local minimum. We use an energy tolerance per particle of $10^{-16}$ in units of $\epsilon$.

In the repulsive, soft particle system, there are several known properties of the forces between overlapping particles: (1) the shape of the distribution of forces~\cite{ohern3,eerd} and (2) the distribution of forces demonstrates a lack of self-averaging~\cite{ohern1,ohern2}.
As for the shape of the distribution, the distribution is not long-tailed.  Recent simulations in two- and three-dimensions have determined the shape of the tail down to normalized force values of $10^{-45}$~\cite{eerd}. In two-dimensions,
the tail is Gaussian. In three-dimensions, the tail falls off slightly more slowly than Gaussian, i.e. $P(f)\sim
e^{-f^{1.7(1)}}$, where $f$ is the magnitude of the force between two overlapping particles.   As for the lack of self-averaging, the distribution of forces takes on one form if the forces in each sample are normalized via the average force per sample or if the forces from all samples are pulled and normalized by the average force from
all samples.  Therefore, one cannot consider a large sample to be comprised of many smaller samples. A lack of
self-averaging is also found in configurations of spin glasses.

While the distribution of forces is a very useful quantity, it does not encode any spatial information. Motivated by the
recent work of Zhou and Dinsmore~\cite{zhou}, we search for spatial correlations in the magnitude of forces. To do this, we measure the angle between
{\it any} two force bonds emanating from a particle in the jammed packing at a particular packing fraction, $\phi$.  We
denote this angle as $\theta$. See the schematic in Figure 1. The probability distribution for $\theta$, $P(\theta)$, is plotted in Figure 1 for all coordination numbers greater than 2 for $\phi=0.841$ with $\lambda=3/2$ and $N=512$.  The lower bound on the coordination number is determined by the principle of local mechanical stability.  For this particular $\phi$ and system size, the average coordination of the jammed configurations, $<z(\phi)>=4.076(2)$. Also, the fraction of jammed configurations is approximately two-thirds.  So we are in the jammed phase.  We should point out that earlier work extrapolated to a critical value of $\phi_c$ in the infinite system limit of approximately $0.842$.  However, there exists a body of recent work suggesting that $\phi_c$ depends on the protocol for obtaining jammed configurations such that ``Point J'' should be modified to ``Segment J''~\cite{chaudhuri,ciamarra,schreck,vagberg}.  Implications of these findings have not been fully fleshed out to date.  Figure 1 also depicts $P(\theta)$ for a particular $z$, or $P_{z=4}(\theta)$.

We then compare $P(\theta)$ with $P(\theta_{lg})$, where $\theta_{lg}$ is the angle between the two
largest force bonds on a particle.  We observe that the probability distribution in the latter case is more heavily biased towards the larger angles. In other words, there exists a suppression of the smaller angles between the two largest force bonds on a particle giving rise to chain-like correlations in the locally largest force bonds. Defining $W=\int_{120^{\circ}}^{180^{\circ}} (P(\theta_{lg})-P(\theta))d\theta$ as a measure of the bias, $W=0.434(1)$ for $\phi=0.841$ and $N=512$.

Let us compare $P(\theta)$ and $P(\theta_{lg})$ with the equivalent distributions for a more ordered packing.  To do this, we use a monodisperse distribution of repulsive, soft particles in two-dimensions at a higher packing fraction. For a hexagonal packing of hard particles in two-dimensions, the packing fraction is equal to approximately 0.907. We choose a slightly larger packing fraction to ensure overlaps. See Figure 2. We observe three dominant peaks at $60^{\circ}$, $120^{\circ}$, and
$180^{\circ}$ degrees indicating a hexagonal packing.  We also note a bias in
$P(\theta_{lg})$ towards the larger angles for the ordered case as well. 

In the disordered case, there are a number of dominant peaks in $P(\theta)$ at $\theta=54^{\circ},60^{\circ},64^{\circ}$, and $70^{\circ}$. There are also some subdominant peaks, many of them separated by intervals of $6^{\circ}$. See, for example, the range of $\theta=102^{\circ}$ to $\theta=114^{\circ}$. These subdominant peaks are due to the bidispersity in particle radii. If the 1.4 ratio is increased, the spacings between the subdominant peaks increases. To see this, consider a particle surrounded by all ``small'' particles. If we replace one of the small particles by a larger particle, the larger particles will push its neighbors away due to
insufficient space and therefore change the contact angle. We can calculate this change given the ratio between the two different
radii in the just-touching case. For a radius ratio of 1.4, the change in contact angle is approximately $5.4^{\circ}$, the interval between the subdominant peaks.

For $N=512$ and $\lambda=3/2$, as $\phi$ is increased from $\phi=0.841$ to $\phi=0.843$, the bias towards the larger angles remains robust.  More specifically, $W$ remains constant.  See Figure 3. For even larger $\phi$, $W$ decreases. For example, $W=0.424(2)$ for $\phi=0.846$. This trend also agrees with the findings of Zhou and
Dinsmore~\cite{zhou} where larger $z$ suppresses chain formation. Since $<z(\phi)>$ increases with $\phi$, this trend is expected. Figure 4 also demonstrates this effect where $P_{z=3,4,5}(\theta_{lg})$ are plotted individually for $\phi=0.843$.  As $z$ is increased from 4 to 5, $W$ decreases from $0.454(2)$ to $0.413(2)$. Finally, the trend of the suppression of smaller contact angles for larger force bonds persists for larger system sizes indicating that the suppression is not a finite system effect. We have also checked that the suppression persists for $\lambda=5/2$ and $\lambda=2$ as well, the more studied cases.

To contrast our results with those of Zhou and Dinsmore~\cite{zhou}, we do not observe a peak in $P_z(\theta_{lg})$ at 180
degrees for all $z$.  Rather the peak in $P_z(\theta_{lg})$ for $z=3$, for example, is closer to $160^{\circ}$. Only for $z=4$, is the peak at $180^{\circ}$. Therefore, the large force bond propagation is not as
``straight'' as in Zhou and Dinsmore~\cite{zhou} simulations where a peak at $180^{\circ}$ is observed for all $z$. Their protocol generates stable configurations by demanding force balance on each particle given that some of the forces on each particle are randomly generated. Zhou and Dinsmore argue that the finding of chain-like correlations is simply a consequence of Newton's third law.  The peaks
in their $P_z(\theta_{lg})$ may be an indication of this. However, since our equivalent distributions are not all peaked at 180 degrees, a more complicated mechanism may be at work.

\begin{figure}
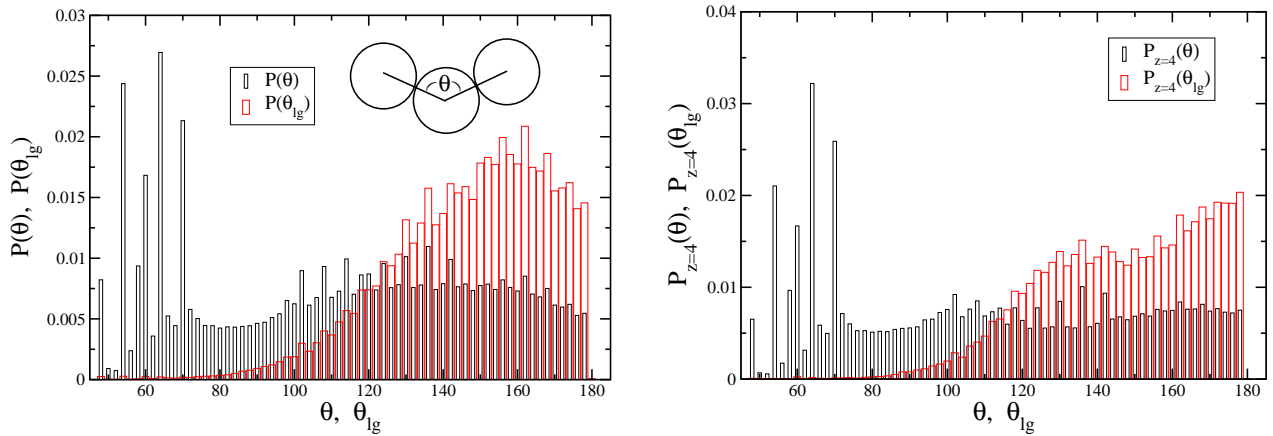

\begin{center}
\includegraphics[width=8cm]{angle.N512.phi0.841.alpha3.2.eps}
\hspace{0.5cm}
\includegraphics[width=8cm]{angle.N512.phi0.841.alpha3.2.z4.eps}
\caption{Left: $P(\theta)$ and $P(\theta_{lg})$ for $\phi=0.841$, $N=512$, and $\lambda=3/2$. Right: $P_{z=4}(\theta)$ and $P_{z=4}(\theta_{lg})$ for the same parameters.  The bin size is $2^{\circ}$.}
\end{center}
\end{figure}
\begin{figure}
\begin{center}
\includegraphics[width=8cm]{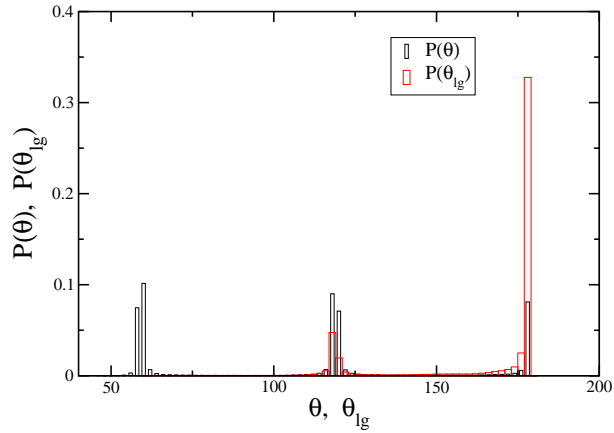}
\caption{$P(\theta)$ and $P(\theta_{lg})$ for $\phi=0.91$, $N=512$, and $\lambda=3/2$ for a monodisperse system.}
\end{center}
\end{figure}
\begin{figure}[b]
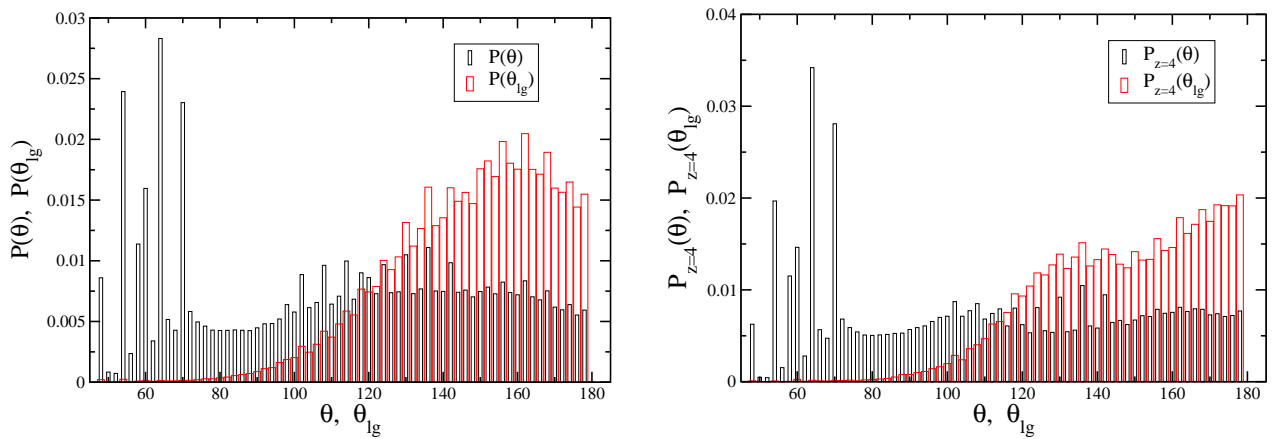

\begin{center}
\vspace{0.25cm}
\includegraphics[width=8cm]{angle.N512.phi0.843.alpha3.2.eps}
\hspace{0.5cm}
\includegraphics[width=8cm]{angle.N512.phi0.843.alpha3.2.z4.eps}
\caption{Left: $P(\theta)$ and $P(\theta_{lg})$ for $\phi=0.843$,
  $N=512$, and $\lambda=3/2$. Right: $P_{z=4}(\theta)$ and $P_{z=4}(\theta_{lg})$ for the same parameters.}
\end{center}
\end{figure}

\begin{figure}
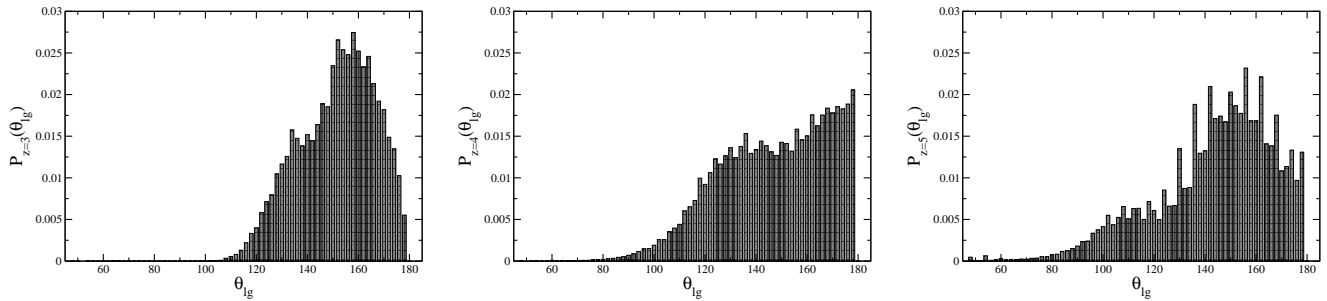

\begin{center}
\includegraphics[width=5.5cm]{angle.N512.phi0.843.alpha3.2.lf.z3.eps}
\hspace{0.25cm}
\includegraphics[width=5.5cm]{angle.N512.phi0.843.alpha3.2.lf.z4.eps}
\hspace{0.25cm}
\includegraphics[width=5.5cm]{angle.N512.phi0.843.alpha3.2.lf.z5.eps}
\caption{Left: $P_{z=3}(\theta_{lg})$ for $\phi=0.843$,
  $N=512$, and $\lambda=3/2$. Middle: $P_{z=4}(\theta_{lg})$ for the same
  parameters. Right: $P_{z=5}(\theta_{lg})$ for the same parameters. }
\end{center}
\end{figure}

Our results suggest a possible spatial correlation in the force bond strength. Given that larger angles between the locally largest forces persist, perhaps locally large force bonds can persist/percolate across the sample.  To test for this, we construct a force
chain by traversing along largest and second largest local force bonds and determining whether the chain of particles spans the system or not. The algorithm for this task is the following: \\

\noindent (1) Identify the largest force bond in the system.\\
(2) Traverse from one of the particles on either side of the largest force bond to the another overlapping particle along the second largest force bond. \\
(3) Traverse to the next overlapping particle along either the largest force bond or the second largest force bond (if the largest force bond has already been traversed).  \\
(4) Repeat (3) until reaching an overlapping particle where both the largest and second largest force bonds have already been traversed. \\
(5) Go back to the largest force bond in the system.  Pick the other particle not initially chosen. If its second largest force bond has not been traversed, repeat steps (4) and (5).  \\

\noindent Note that the largest force bond
of one particle is not necessarily the largest force bond of the other particle associated with the bond. This algorithm constructs the largest force chain in the sample.  We also construct the weakest force chain in the sample by replacing largest with smallest, etc. Finally, we also study force chains beginning from any force bond in the system.

In Figure 5 we present an example of the largest force chain.  Loops make up some fraction of the ``chain''.  This is
due to fluctuations in the forces. The larger force bonds exhibit chain-like structures, while smaller force bonds form local loops.  We have checked this in the simulations by constructing the equivalent weakest force chain. Since the largest local force bond between two particles may be one of the smaller force bonds, as compared to the rest of the force bonds in the system, chains can end in loops.
Note that, at least for this example, the largest force chain spans the system in the vertical direction.

How typical is this spanning for the largest force chains?  Figure 6 shows the probability of spanning in either direction, $P_s$, as a function of $N$ for a particular $\phi$. We see that $P_s$ decreases with
increasing $N$.  While for small $N$, $P_s$ appears to decay linearly with
$N$, the larger particle number data makes this candidate function less
likely.  We also observe that $P_s$
increases as $\phi\rightarrow \phi_c$, particularly for the largest particle
number data, which is consistent with the increase in small angle suppression
as $\phi\rightarrow \phi_c$ for the locally largest forces. As for the infinite system limit, it may be that the conditions
to generate spanning chains need to be relaxed to generate spanning chains
with nonzero probability.  Also, if we start from any force bond in the
system and generate a force chain moving along the locally largest two force
bonds for each particle, the probability for spanning decreases. To be specific, for $\phi=0.841$, $N=1024$, and $\lambda=3/2$, $P_s\approx0.563$ starting at any force bond as compared to the largest force chain spanning of probability, where $P_s\approx 0.688$. 

At this point, we must comment on the relation of our results to those of Makse and collaborators~\cite{makse} performing dynamical simulations of deformable grains.  They construct spanning force chains by starting with a grain at one end of the system and traversing the maximum force bond at every particle such that they reach the other side of the system. They find fewer spanning force chains closer to the transition than further away. They argue that this observation is due to an increasingly homogeneous distribution of forces far away from the transition.  We should also contrast our results with work by Ostojic and collaborators~\cite{ostojic}.  They study the spatial extent of force bonds exceeding some threshold force.  The threshold force is lowered until the force bonds exceeding the threshold force span the system. They observe a percolation transition of a new universality class.  We, instead, investigate the
local, largest forces.  Also, we do not have spanning as a criterion, but rather as an
``afterthought''.  

To better understand the spatial properties of the largest force chain, we measure its fractal dimension.  Presumably, the fractal dimension is unity, though one should check this. To do so, we count the number of particles participating in the largest force chain.  To relate this
number with a length, in two-dimensions, $L\sim \sqrt{N}$. Figure 7 plots the average number of particles participating
in the largest force chain, $N_{FC}$, as a function of system size, $N$. On the log-log scale, there is some
curvature to the data. If we assume a fractal form, for $\lambda=3/2$ and $\phi=0.843$, we measure a fractal dimension of
1.10(5)--very close to unity.  Comparing our largest force chain data to the
Ostojic formulation~\cite{ostojic}, we measure the number of particles participating in the largest force spanning cluster, $N_{LF}$. The fractal dimension of the spanning cluster of largest forces is 1.62(2).  We then measure the number of particles participating in the smallest force spanning cluster, $N_{SF}$, where we replace ``exceeding a threshold force'' in the Ostojic formulation~\cite{ostojic} with ``below a threshold force'', the fractal dimension is 1.68(2). The fractal dimension of the spanning cluster at the ordinary percolation transition is about 1.89~\cite{percolation}, which is somewhat larger than the value measured for the spanning cluster of largest forces (and smallest forces). This discrepancy may be due to the chain-like correlations in the largest, local force bonds, though one cannot discount the possibility of an eventual crossover to ordinary percolation. 

\begin{figure}
\begin{center}
\includegraphics[width=7cm]{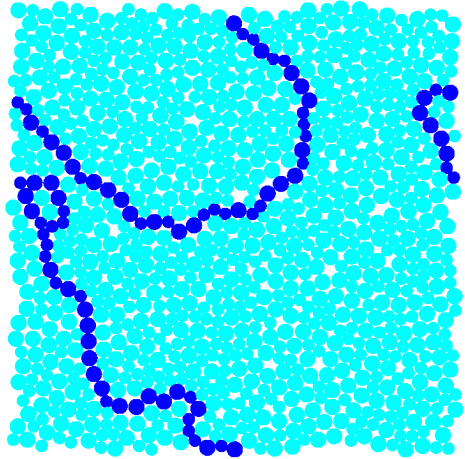}
\vspace{-3cm}
\caption{Jammed configuration for $\phi=0.841$, $N=1024$. The particles participating along the largest force chain (dark blue) as
distinguished from the other particles (light blue).}
\end{center}
\end{figure}

\begin{figure}
\begin{center}
\vspace{0.3cm}
\includegraphics[width=8cm]{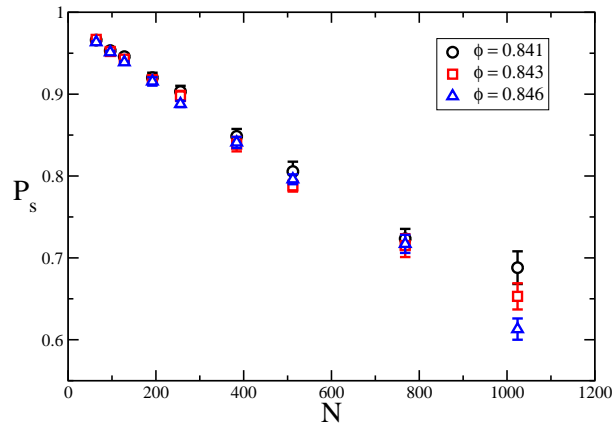}
\caption{The probability of spanning in either direction, $P_s$, for the largest force chain.}
\end{center}
\end{figure}
\begin{figure}
\begin{center}
\includegraphics[width=8cm]{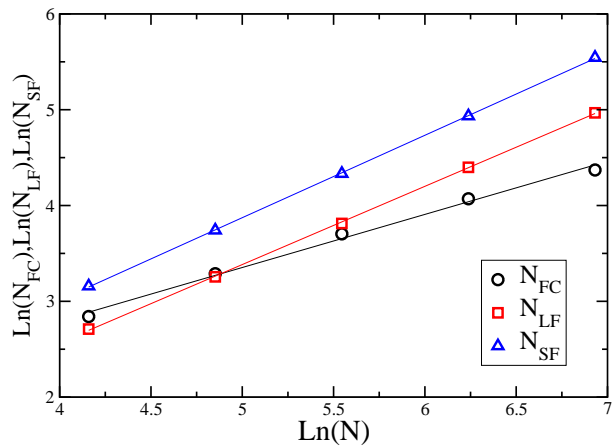}
\caption{Log-log plot of the average number of particles in the largest force chain that span, $N_{FC}$, the average number of particles in the largest force spanning cluster, $N_{LF}$, and the average number of particles in the smallest force spanning cluster, $N_{SF}$, as a function of $N$. The error bars are smaller than the symbols. The straight lines are fits to the data. Here, $\phi=0.843$ and $\lambda=3/2$. }
\end{center}
\end{figure}
\section{Coordination number and spins}

The forces are not the only information encoded in the force network.  The coordination number is another piece
of information.  Figure 8 is the same jammed configuration as depicted in Figure 5. Each color now represents a
different coordination number.  If the color of each vertex, a particle, shared by an edge, a contact, were of a different
color, then this graph would represent a proper vertex coloring.  However, note that there are some vertices sharing an
edge that do have the same color.  So Figure 5 does not represent a proper vertex coloring.

But let us, for the moment, address proper vertex colorings. A proper vertex coloring using at most $p$ colors exists if there exists a zero energy solution of zero-temperature antiferromagnetic Potts model with $p$ states. In other words, the $p$-coloring problem maps to a zero-temperature antiferromagnetic Potts model with $p$ states. Now, if one were to consider the jamming problem, the number of possible colors participating in the jammed clusters is 5 ($z=3$ through $z=7$, for this particular two-dimensional bidisperse system) which leads to a $p=5$ state Potts model.  At $\phi=\phi_c$, the isostatic condition would bias
one of the colors---for example, the color
associated with coordination number four dominates for two-dimensional frictionless discs.  This bias can be encoded via a magnetic field. The isostatic condition also imposes other weights on the
other colors as well such that the average coordination number is four.  In particular, recent work shows that at the jamming transition in two-dimensions, the coordination numbers range from 3 to 5 such that $z=4$ particles make up about half the system and $z=3,5$ particles make up the other half~\cite{henkes3}. As the packing fraction increases, the weights for each color
changes. For a hexagonal packing, all couplings are ferromagnetic.  

Both the antiferromagetism and the competing magnetic fields (to account for the average coordination number) contribute to frustration of the spin system. Moreover, there exists randomness in the system due to (1) the fact that not all particles are participating in the jammed configuration and (2) the randomness of the forces. The combination of frustration and disorder should lead to a spin glass phase~\cite{edwards,sherrington}--a spin glass
phase with no quenched disorder.  In fact, the couplings and dilutions (particles not participating in the jammed configurations) are dynamically generated. However, the lack of
self-averaging in the forces is characteristic of quenched disordered system.
Indeed, there are systems with no disorder that
can be mapped to quenched disorder and there are systems with unquenched disorder that can also be mapped to quenched
disorder.  For the former, consider an Ising spin system whose interaction fluctuates in sign with distance between sites~\cite{marinari}.
For the latter, consider a liquid of hard spheres as described by hypernetted chain equations~\cite{zamponi}.

The potential for a jamming-spin glass correspondence provides motivation to analyze the jamming system in terms of a spin glass.  If we compute the ratio of antiferromagnetic couplings (different coordination numbers (colors) between contacts) to ferromagnetic couplings (same coordination numbers (colors) between contacts), for $\phi=0.841$, $N=1024$, for example, the ratio is approximately 2.5.  We could ignore the random variation in the magnitude of the forces (interactions between spins) and set the coupling to be a constant using the measured ratio of antiferromagnetic spins to ferromagnetic spins to simulate a two-dimensional $p=5$ Potts glass system at zero-temperature.  While there exists a number of ground state algorithms for two-dimensional Ising spin glasses~\cite{barahona,thomas,pardella}, we do not know of an exact ground state algorithm for the 5-state Potts glass and so we leave this avenue for future work.  

Instead, to further the possible connection between jammed solids and spin glasses, we compute the spin glass equivalent of bond chaos in the repulsive, soft particle system~\cite{bray1}. Chaos in mean-field spin glasses is presumably due to the fact that the spin glass phase is a marginal phase such that a perturbation in the energy of the system is sufficient to alter the weights of different equilibrium configurations~\cite{ritort}. More specifically, in mean-field, the Hessian associated with Parisi's ansatz for the structure of the matrix order parameter contains all non-negative eigenvalues~\cite{dominicis}.  In finite-dimensions, the marginality is presumably due to the slowly decaying correlation functions~\cite{neynifle}. It is interesting to note that jammed solids at the jamming transition are marginally rigid and, therefore, may also exhibit similar chaotic features. 

So, we use this notion of chaos in spin glasses to look for a quantitative measure of chaos in the jamming system. To do this, we first assign a list of random numbers as the initial position of particles in a system and use conjugate gradient algorithm to generate a jammed state. Next, we use the same initial positions and perturb them by a given strength, then apply the same conjugate gradient algorithm to generate another jammed state. Specifically, for particle $i$ with initial position $(x_i,y_i)$ in the unperturbed system, the corresponding particle $i$ in the perturbed system has an initial position of
\begin{eqnarray}
x_i'&=&x_i+\delta\,c_i \nonumber \\
y_i'&=&y_i+\delta\,d_i,
\end{eqnarray}
where $\delta$ is the magnitude of perturbation strength, and $c_i$ and $d_i$ are randomly generated numbers chosen from the same distribution used to generate $x_i$ and $y_i$.

If the two sets of initial positions both lead to a jammed state, we calculate the overlap between the two states as defined by
\begin{equation}
r(\delta,N,\phi)=<\frac{\sum_{j=1}^{M}\cos^2(S_j-S_j')}{M^2}>,
\end{equation}
where $S_j$ is the spin state of the unperturbed system (using the Domb representation), $S_j'$ is the spin state of the perturbed system, and $M$ is the number of spins in both systems with $z>2$, which grows with $N$. The brackets denote ensemble averaging. For $\delta=0$, $r=1$.  The system is chaotic when
\begin{equation}
\lim_{\delta\to 0}\lim_{N \to \infty}r(\delta,N,\phi)<1.
\end{equation}
It may be that the limit of $\phi\rightarrow \phi_c$ may also have to be taken to observe chaos given that only at the jamming transition is the system marginally rigid.

In Figure 9 we present results for $r$ as a function of $N$ for different values of $\delta$ and $\phi$.  For fixed $\phi$, we observe that $r$ initially decreases with increasing $N$ and then begins to level off (though one cannot rule out an small increase in $r$ for the largest $N$ studied). Moreover, $r$ increases with decreasing $\delta$ for fixed $N$, as expected. Since it is not clear that $r$ reaches a limit as a function of $N$ for fixed $\delta$, we cannot easily extrapolate to the zero perturbation limit. However, it is clear from the data that as $\phi$ decreases towards unjamming, $r$ decreases, indicating that as the jamming transition is approached from above, the system is becoming increasingly chaotic.

\begin{figure}
\begin{center}
\includegraphics[width=7cm]{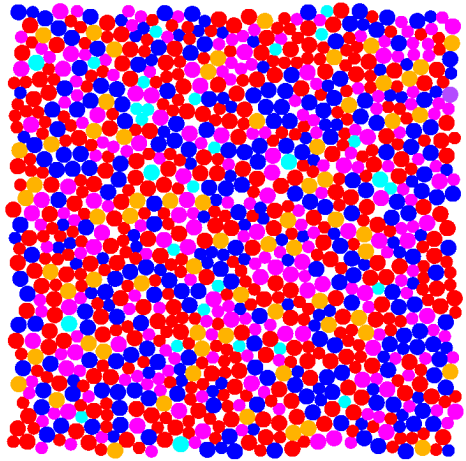}
\vspace{-3cm}
\caption{Jammed configuration (same as above) colored via coordination number.
Light blue denotes $z=0$, magenta denotes $z=3$, red denotes $z=4$, blue depicts $z=5$, orange depicts $z=6$, and purple denotes $z=7$ (possible for 1.4 diameter ratio bidisperse system).}
\end{center}
\end{figure}

\begin{figure}
\begin{center}
\includegraphics[width=8cm]{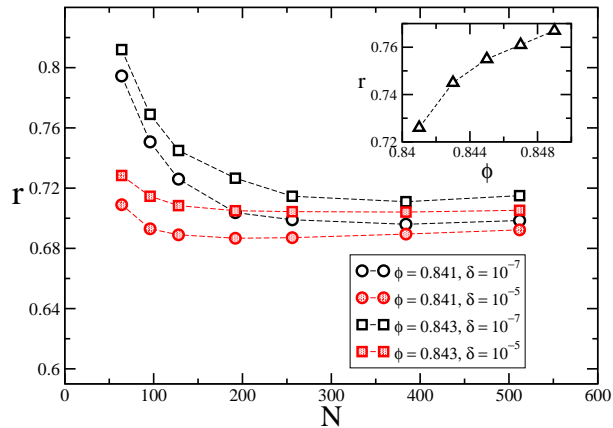}
\caption{Plot of $r$ as a function of $N$ for several $\delta$s.  Inset: Plot of $r$ for $N=128$ as a function of $\delta$.  The error bars are smaller than the symbols.}
\end{center}
\end{figure}

\section{Jammed solids and spin glasses}

Let us summarize our numerical findings:\\
\\
(0) There is a suppression of the smaller angles subtending locally larger force bonds. This suppression increases as $\phi_c$ is approached from above. \\
(1) The probability for spanning of the largest force chain increases as $\phi_c$ is approached from above.\\
(2) The coordination number-spin state mapping suggests that the jammed solid becomes increasingly chaotic as
$\phi_c$ is approached from above.\\

Keeping the above findings in mind, we recall one of the traditional models of an amorphous solid via a randomly dilute network of springs with probability $p$, the model otherwise known as rigidity percolation~\cite{feng}. While the nature of the rigidity percolation transition remains contentious~\cite{jacobs1,moukarzel}, an important concept has emerged in terms of constraint
counting, a concept initially presented by Maxwell back in 1864~\cite{maxwell}.  Below the rigidity percolation threshold,
there are only underconstrained bonds, above the threshold, there exists overconstrained bonds.  At the transition, on
average, the system is not overconstrained or underconstrained---it is isostatic. This condition should then determine the location of the
transition. For central force interactions, $p_r=2d/z$, where $d$ is the dimension of the system and $p_r$ is the occupation probability above which the system is rigid. For the
triangular lattice, $p_r=2/3$. Numerical studies find a result close to this estimate.  However, the Maxwell argument
does not take into account redundant bonds and the possibility of a fractal set of overconstrained bonds participating
in a rigid cluster.

Jacobs and Thorpe have numerically extended Maxwell's argument by identifying those overconstrained bonds via an algorithm
implemented on a two-dimensional bar-joint network~\cite{jacobs2}, which makes uses of a theorem from Laman~\cite{laman}.  They demonstrate that at the transition, the overconstrained bars make up a fractal subset of the bonds. As more bars are added, the
fraction of overconstrained bars increases such that they make up a finite fraction of the bars. It turns out that the
fraction of overconstrained bars can be viewed as the order parameter for the rigidity percolation transition.

Indeed, it would be interesting to extend some of these ideas of rigidity percolation more concretely to the repulsive, 
soft particle system, such as trying to identify the overconstrained bonds via use of the pebble game algorithm.  However, use of the pebble algorithm takes into account only fixed connectivity information and it does not take into account the force information.   Of course, there have been several recent works drawing a more intimate connection between rigidity percolation and jamming. Ellenbroek and collaborators map the jammed solid to a network of stretched springs~\cite{ellenbroek3}.  Study of the stretched spring network demonstrates anomalous scaling of the bulk modulus as opposed to the shear modulus.  Even more recent work investigates a square lattice occupied by springs~\cite{mao}.  Such a lattice is isostatic with a frequency-independent density of states.  As next-nearest springs are occupied, the system undergoes a rigidity percolation transition for an infinitesimal occupation of next-nearest neighbor springs.  This system exhibits the kind of transition proposed by the phenomenological framework put forth by Wyart and collaborators.  There exists a clear lengthscale below which the system behaves isostatically and above which it does not, which is not so readily apparent in the disordered packing, at least in terms of a direct measurement. 

While working within the more traditional spring framework is certainly useful, coupling the coordination number-spin state mapping with the force information via interaction strengths between the spins and the dilution information provides one with an alternate description of a jammed solid---a spin description to be compared to the traditional framework. For example, overconstrained bonds in rigidity percolation may correspond to frozen spins---spins that take on the same state in every configuration---in the spin picture. From the numerical information obtained so far, there are more antiferromagnetic interactions than ferromagnetic interactions. As for the magnitude of the interactions between the spins, recall the chain-like correlations observed in the locally larger force bonds.  This information can be also be encoded into a Potts spin Hamiltonian via chain-like correlations in the spin interaction strengths.

How does the Potts spin Hamiltonian change with increasing packing fraction? The amount of dilution decreases. Moreover, the ratio of antiferromagnetic to ferromagnetic interactions also decreases. What about the force information? We observe that the probability of spanning increases with decreasing $\phi$ for the largest system sizes. Based on this observation, the chain-like correlations in the spin interaction strengths decrease as the packing fraction increases. All of these properties can be encoded into a Potts spin Hamiltonian. 

Given the interplay of frustration and randomness, we conjecture that a Potts spin analog of the jammed solid should exhibit spin glass behavior. A possible spin glass/jammed solid correspondence may help to explain the ordinary elasticity findings of Ellenbroek and collaborators at all lengthscales when measuring the system's averaged response to inflation of a central particle~\cite{ellenbroek1,ellenbroek2}.  For both the paramagnet and the spin glass, the average magnetization is zero, i.e. one cannot distinguish between the two phases.  To observe the spin glass phase, one must measure higher order quantities such the Edwards-Anderson order parameter. Ellenbroek and collaborators did observe a deviation from ordinary elasticity when measuring fluctuations in the forces averaged over all
contacts within a distance from the perturbation~\cite{ellenbroek1,ellenbroek2}. The lengthscale beyond which the fluctuations level off may correspond to a finite-dimensional spin glass phase, which exhibits very different properties than a mean-field spin glass. The replica, or mean-field, scenario~\cite{replica,spinglassbook} gives evidence for infinite number of pure thermodynamic states,
while the low-dimensional, or droplet scenario~\cite{fisher1,fisher2}, at least in the Ising case, argues for two groundstates, the same ones occurring in the Ising
ferromagnet---all spins up or all spins down.    In the droplet picture, an applied magnetic field is a relevant perturbation driving the system to a ferromagnetic phase, while in the replica scenario, an infinitesimally applied magnetic field does not destroy the spin glass phase. How such a lengthscale separating mean-field and low-dimensional behavior could come about, we propose in the following subsection.

\subsection{Simple example: Long-range, one-dimensional Potts glass}

While one would ultimately like to simulate a zero-temperature, two-dimensional Potts-glass with chain-like correlations in the couplings, for the time being we make use of a simple example, namely, the one-dimensional, long-range interacting $p$-state Potts glass with Gaussian distributed quenched bond randomness.  There exists previous analysis of the one-dimensional, long-range interacting Ising spin glass. In light of our coordination number-spin state mapping, we extend these results to the Potts glass case.  

Why study a model with long-range interactions? The chain-like correlations in the forces may call for an effective theory centered on these chains
with the non-force chain bonds providing for long-range forces along the chain. In fact, the existence of these chain-like correlations is reminiscent of a picture proposed earlier by Cates and collaborators~\cite{cates} where the jammed solid is comprised of linear force chains and a sea of spectator particles modeled as an incompressible solvent.  Instead of considering the force chain as a linear object supporting loads only along its own axis, consider the force chain as a polymer---a polymer embedded in solution, i.e. the other particles.  Just as the hydrodynamics of the solution couple one part of the polymer 
with a distant part, resulting in long-range interactions along the polymer, the force chain would also experience long-range interactions~\cite{zimm}. We note that the polymer-in-solution analogy has its shortcomings.  A more accurate analogy might be a polymer embedded in another polymer-type medium. Also, resumably the particles participating in the largest force chain would no longer
be part of the largest force chain upon shearing or other perturbations.  In other words, there is particle conversion
between the force chain and non-force chain particles. 

How would this effective force chain framework depend on the packing fraction? 
At the onset of jamming, presumably there exists an isotropic, fractal network of larger force spanning chains. As the packing fraction is increased, the larger force chains become shorter such that
eventually there is no distinction between the chains and the other particles since the definition of a force chain would have to be relaxed in order to achieve spanning.  So the lengthscale of the chain lattice shrinks as packing fraction is increased and the long-ranged interactions become screened out just in the case of many polymers in solution. At a particular packing fraction; while for lengthscales smaller than ``chain'' lattice, the system behaves
long-ranged, or mean-field-like; for lengthscales larger than ``chain'' lattice, the long-range interactions are
screened out and the system behaves as a low-dimensional system. Therefore, within one system, one can interpolate between mean-field and finite-dimensions due to the correlated heterogeneity in the forces/interactions.

While a precise formulation of an 
effective force chain theory is still lacking, we discuss the possible ramifications of long-range interactions in a conventional
spin glass system. We begin with the $p$-state Potts glass Hamiltonian in $d$-dimensions, 
\be
H=-\sum_{i,j}\frac{J_{ij}}{|i-j|^\sigma}\sum_{a=1}^{p-1}S_{ia}S_{ja},
\ee
where ${\bf S}_i$ is a $p$-state Potts spin in the simplex representation~\cite{zia} at site $i$ and $\sigma$ denotes the interaction range. The distribution of the coupling constant $J_{ij}$ is given by
\be
P(J_{ij})=\frac{1}{\sqrt{2\pi J^2}}e^{-(J_{ij}-J_0)^2/2J^2}.\nonumber
\ee

The replica trick is applied to the Hamiltonian to compute the free energy averaged over disorder by considering $n$ identical replicas of the original system. The averaged free energy, $F$, is then given by 

\be F=-k_BT\lim_{n\rightarrow 0}{\frac{\overline{Z^n}-1}{n}},  \ee
where
\be \overline{Z^n}=\int(\prod_{i,j}{P(J_{ij})dJ_{ij}})
Tr_{\{S_{i}^{\alpha}\}}\exp{\left\{\sum_{\alpha=1}^n
\sum_{i,j}\frac{J_{ij}}{k_BT|i-j|^\sigma}\sum_{a=1}^{p-1}S_{ia}^{\alpha}S_{ja}^{\alpha}\right\}}.\nonumber
\ee

After integrating over the disorder and applying the Hubbard-Stratonovich transformation, 
\be
\overline{Z^n}=\int\prod_{\alpha\neq\beta\atop{a,b}}dQ_{i,ab}^{\alpha\beta}\exp\left\{-\frac{1}{2}K_{ij}^{-1}Q_{i,ab}^{\alpha\beta}Q_{j,ab}^{\alpha\beta}\right\}Tr_{\{S_i^{\alpha}\}}\exp\left\{S^\alpha_{ia}S^\beta_{ib}Q^{\alpha\beta}_{i,ab}\right\},
\ee
where the matrix $K$ is defined by
\be
K_{ij}=\frac{J^2}{(k_BT)^2}\frac{1}{|i-j|^{2\sigma}}
\ee
and $\frac{J_0}{J}<\frac{4-p}{2}$ with $J=k_BT_g$ such that we set the ferromagnetic order parameter to zero. 

To perform the trace over the spins, we recall that the simplex representation of the $p$-state Potts model requires

\begin{eqnarray}
\sum_s e_a^s e_b^s &=& p\delta_{ab},\nonumber \\
\sum_a e_a^s e_a^{s'} &=& p\delta_{ss'}-1,  \\
\sum_s e_a^s  &=& 0, \nonumber
\end{eqnarray}
where in the replicated partition function each $S_{ia}$ is represented by one of the $e_a$'s. We also define $v_{abc}=\sum_s{e_a^s e_b^s e_c^s}$ for convenience~\cite{cwilich}. We expand the second exponential term in Eq. 7 to order $Q^3$, and after computing the trace and re-exponentiation, we obtain 
\begin{eqnarray}
Tr_{\{S_i^{\alpha}\}}\exp\left\{S^\alpha_{ia}S^\beta_{ib}Q^{\alpha\beta}_{i,ab}\right\}=\frac{1}{4}\sum_{\alpha\neq\beta}(Q_{i,ab}^{\alpha\beta})^2+\frac{1}{6}\sum_{\alpha\neq\beta\neq\gamma}Q_{i,ab}^{\alpha\beta}Q_{i,bc}^{\beta\gamma}Q_{i,ca}^{\gamma\alpha}
\nonumber \\ 
+\frac{1}{12}\sum_{\alpha\neq\beta}\frac{v_{abc}v_{def}}{p^2}
Q_{i,ad}^{\alpha\beta}Q_{i,be}^{\alpha\beta}Q_{i,cf}^{\alpha\beta}+\mathcal{O}((Q_{i,ab}^{\alpha\beta})^4).
\end{eqnarray}
After taking the continuum limit, we rewrite $\overline{Z^n}$ in momentum space as  
\begin{equation}
\overline{Z^n}=\int\prod_{(\alpha,\beta) \atop {a,b}}dQ_{ab}^{\alpha\beta}(\vec{q})\exp(-\int d^d\vec{q} \,H\{Q_{ab}^{\alpha\beta}(\vec{q})\})
\end{equation}
with 
\begin{eqnarray}
H=&&\frac{1}{4}\int_q\sum_{\alpha\neq\beta}(r+q^{2\sigma-d})Q_{ab}^{\alpha\beta}(\vec{q})Q_{ab}^{\alpha\beta}(-\vec{q})-\frac{1}{6}v\int_{q_1q_2q_3}\sum_{\alpha\neq\beta\neq\gamma}(2\pi)^d\delta(\vec{q}_1+\vec{q}_2+\vec{q}_3)Q_{ab}^{\alpha\beta}(\vec{q}_1)Q_{bc}^{\beta\gamma}(\vec{q}_2)Q_{ca}^{\gamma\alpha}(\vec{q}_3)\nonumber\\ &&-\frac{1}{2}u\int_{q_1q_2q_3}\sum_{\alpha\neq\beta}(2\pi)^d\delta(\vec{q}_1+\vec{q}_2+\vec{q}_3)Q_{ad}^{\alpha\beta}(\vec{q}_1)Q_{be}^{\alpha\beta}(\vec{q}_2)Q_{cf}^{\alpha\beta}(\vec{q}_3)v_{abc}^{\alpha}v_{def}^{\beta},
\end{eqnarray}
where $r$, $u$, and $v$ are the coupling constants and $\int_q=\int \frac{d^d\vec{q}}{(2\pi)^d}$. We have omitted a $q^2Q_{ab}^{\alpha\beta}(\vec{q})Q_{ab}^{\alpha\beta}(-\vec{q})$ term. 
 
Performing the momentum shell renormalization group with one-loop corrections using the usual notation~\cite{harris}, we obtain 
\begin{eqnarray}
r'&=&\zeta^2b^{-d}\left(r-(-2(p-1)v^2+18p^4(p-2)^2u^2)C(b)\right)\nonumber\\
v'&=&\zeta^3b^{-2d}\left(v+(-3p+4)v^3D(b)\right)\nonumber\\
u'&=&\zeta^3b^{-2d}\left(u+36u^3p^4(p-3)^2D(b)\right)\nonumber\\
\zeta&=&b^{1+d/2-\eta/2}\nonumber\\
\end{eqnarray}
where $C(b)=\int^{1}_{1/b}\frac{d^d\vec{q}}{(2\pi)^d(q^{(2\sigma-d)}+r)^2}$ and $D(b)=\int^{1}_{1/b}\frac{d^d\vec{q}}{(2\pi)^d q^{3(2\sigma-d)}}$. Since the renormalization group does not generate new long-range terms, $\eta=2+d-2\sigma$ to all orders. Therefore, for $\sigma<\frac{2d}{3}$, both cubic couplings are irrelevant and the critical behaviour is mean-field~\cite{gross} with Gaussian fluctuations. For $\sigma>\frac{2d}{3}$, the cubic terms become relevant and different critical behaviour is observed. 

The modified value of $\eta=2+d-2\sigma$ is a general feature of long-range interactions. The $v$ and $u$ terms are
irrelevant at $\sigma<2/3$ in the one-dimensional case. Thus, the system behaves mean-field-like when
$1/2<\sigma<2/3$, where the lower bound $\sigma<1/2$ assures the convergence of free energy. This result is a simple
generalization of the previous $p=2$ case~\cite{kotliar,moore} and the short-range case~\cite{chang}.  Modifying the range of the interaction corresponds to modifying the
effective spatial dimension of the system from low-dimensional to mean-field. This property has been numerically verified in the $p=2$ case and the 3-spin spin glass~\cite{katzgraber1,katzgraber2}. In the jammed solid, the interaction range is presumably tuned by the presence of other spanning force chains (Potts chains). On lengthscales smaller than the chain lattice, the long-range interaction dominates and above this lengthscale, the short-range interaction dominates.

\section{Discussion}
 The existence of force chains have been demonstrated both experimentally~\cite{howell} and numerically~\cite{makse}.  We numerically demonstrate the existence of chain-like correlations in the locally large forces in the repulsive, soft particle system. This finding may call for an updated effective force chain description with the non-force chain particles providing for long-range interactions along the force chains as the basis of the description. One can test for these long-range interactions by perturbing a force chain locally and observing the response further along the force chain.

Experiments by Majumdar and Behringer~\cite{majumdar} have shown an ordering of force chains in the presence of shear. In the absence of shear, the network of locally large force chains is isotropic. In the presence of shear, the force chains align along the shear direction. A reordering of the force chain network structure via non-affine displacements such that the system is inherently anisotropic may account for the anomalous shear modulus exponent found in the repulsive, soft particle system. One can test for this reordering in the repulsive, soft particle system by checking for spanning of the locally large forces in the direction along and perpendicular to the shear separately. 

We also propose a coordination number-spin state mapping. With this mapping, we find that the repulsive, soft particle system becomes increasingly chaotic as the the jamming transition is approached from above.  More testing of this mapping is needed.  For example, one can investigate the response of the system to shear and see how the coordination number changes throughout the system.  A changing coordination number corresponds to a changing spin state.  It may be that the framework of random spin wave theory (or some alteration thereof) can be used to understand the response of the jammed solid to various perturbations.

Based on the fact that jamming and glass transition share some commonalities, via transitivity, one suspects that the spin glass transition should be related to the jamming transition. A particle (graph) coloring with non-trivial biasing to enforce the isostaticity condition at the onset of jamming may provide us with the third missing link.  Whether or not the equivalent spin glass is mean-field or low-dimensional remains to be seen.  Of course, both the marginality and the massless, or replicon, modes found in mean-field spin glasses make for a tantalizing match~\cite{bray2,bray3}. A link between jammed solids and spin glasses also provides a connection to constraint satisfaction problems.  This connection has recently been explored by Mailman and Chakraborty~\cite{mailman} and also by Krazkala and Kurchan~\cite{kurchan1,kurchan2}. Constraint satisfaction problems are at the heart of attempting to understand what makes problems solvable. 

While the phenomenological isostatic framework provides insight into the jamming transition, its success hinges on the measured exponents for system sizes ranging up to approximately 10000 particles currently.  History teaches us that the reliance on numerical results for support may ultimately reveal oversimplified assumptions~\cite{toninelli}. Hence, the importance of constructing well-defined, calculable toy models of jamming transition cannot be underestimated. Such is the case for $k$-core percolation, the simplest model of rigidity percolation~\cite{schwarz}.  If one allows the $k$-core percolation transition to take place as soon as a non-zero spanning $k$-core cluster is allowed, i.e. minimal rigidity, then the mean-field exponents are presumably in the same category as those measured for the finite-dimensional repulsive, soft sphere system. Note that minimal rigidity here does not translate to an average bond occupation of four per site.  While the $k$-core constraint is indeed a local one, it contains the nonlocal feature of rigidity percolation--that the removal of a single bond can trigger the removal of an entire occupied cluster.  Other spin systems, such as the dilute 3-state Potts antiferromagnet on the triangular lattice, also exhibit a similar nonlocal property~\cite{adler}. There exists other work towards calculable models of jammed solids that should be pursued further. For example, see Refs.~\cite{socolar,blumenfeld,henkes4}. 

In closing, given the connection between spin systems and repulsive, soft particle systems, we expect that construction and study of new spin models with constraints inspired by jamming will pave the way for further insights into jammed solids.  Should one of these candidate spin models exhibit spin glass properties, we may come full circle to a percolation view of jamming as has been developed for spin glasses~\cite{machta}. 

JMS would like to acknowledge financial support from DMR-0645373.\\

\end{document}